\documentclass[aip,apl,reprint,numerical]{revtex4-1}
\usepackage{latexsym,amssymb}
\usepackage[latin1]{inputenc}
\usepackage{graphicx}
\usepackage{color}
\begin{document}
\title{Strain determination in the Si channel above a single SiGe island inside a field effect transistor using nanobeam x-ray diffraction}
\author{N. Hrauda}
\email{Nina.Hrauda@jku.at}
\author{J. J. Zhang}
\author{E. Wintersberger}
\author{T. Etzelstorfer}
\author{J. Stangl}
\author{G. Bauer}
\affiliation{Institute for Semiconductor Physics, Johannes Kepler University Linz, Altenberger Str. 69, 4040 Linz, Austria}
\author{D. Carbone}
\affiliation{European Synchotron Radiation Facility (ESRF), B.P. 220, F-38043 Grenoble Cedex, France}
\author{C. Biasotto}
\author{V. Jovanovic}
\author{L. K. Nanver}
\affiliation{DIMES, TU Delft, Feldmannweg 17, 2628CT, The Netherlands}
\author{J. Moers}
\author{D. Gr\"{u}tzmacher}
\affiliation{Halbleiter-Nanoelektronik (IBN-1) Forschungszentrum J\"{u}lich GmbH,D-52425 J\"{u}lich}

\begin{abstract}
SiGe islands are used to induce tensile strain in the Si channel of Field Effect Transistors to achieve larger transconductance and higher current driveabilities. We report on x-ray diffraction experiments on a single fully-processed and functional device with a TiN+Al gate stack and source, gate, and drain contacts in place. The strain fields in the Si channel were explored using an x-ray beam focused to 400~nm diameter combined with finite element simulations. A maximum in-plane tensile strain of about 1\% in the Si channel was found, which is by a factor of three to four higher than achievable for dislocation-free tensile strained Si in state-of-the-art devices. 
\end{abstract}

\maketitle 

Tensile strained silicon channels are playing an important role in the design of field-effect transistors (FET) with enhanced electron mobility. \cite{Schmidt&Eberl2001, Ieong2004} Already existing n-channel devices that utilize the stress applied by the gate stack itself\cite{Ieong2004}, buried planar SiGe structures etched out of a layer \cite{Donaton2006}, or Si$_{1-y}$C$_y$ in the source and drain regions\cite{Ang2007, Ang2008} exhibit a scaling problem which needs to be addressed in the coming years according to the International Technology Roadmap for Semiconductors. \cite{ITRS2009} We investigate a different approach, the so called ''dotFET'': epitaxially grown islands, structures resulting from strain relaxation by the lattice mismatch of 4.2\% between Silicon and Germanium. The advantage of this technique is the ability to reach higher Ge contents within the stressor itself without defect formation, whereby  more tensile strain can be induced in the silicon capping layer above. In the dotFET devices the size of the active channel exactly matches the size of one buried island. A regular array of SiGe islands is required to make the concept compatible with Si process technology.
Our method of choice to investigate strain in such structures is x-ray diffraction (XRD). In the past we have reported on investigations involving XRD measurements on whole arrays of SiGe islands, both uncapped and capped \cite{Hrauda2009a, Hrauda2009b}. Due to the uniformity of the islands grown on patterned Si substrates, we were able to use beams with a size of several hundred $\mu$m or even mm. 

In this Letter the focus is on the determination of the strain fields in a fully-processed n-channel FET, the electronic function of which was investigated before the XRD experiments, with only a single dot of the whole array integrated into this device. The tensile strained Si capping layer above it forms the channel between source and drain. Therefore, unlike previous sample investigations by XRD, it becomes mandatory to perform the diffraction experiment only on one single island, and additionally it is required that specifically the island below the transistor gate must be aligned into the x-ray beam. An additional challenge for the success of this experiment was the small scattering volume provided by that island, having a diameter of 220 nm and a height of 45 nm. We show that we were able to find a specific island by means of scanning x-ray diffraction (SDX)\cite{Mocuta2008} techniques and record enough data to base our strain simulations on.
\begin{figure}[tb]
\centering
\includegraphics[width=8cm]{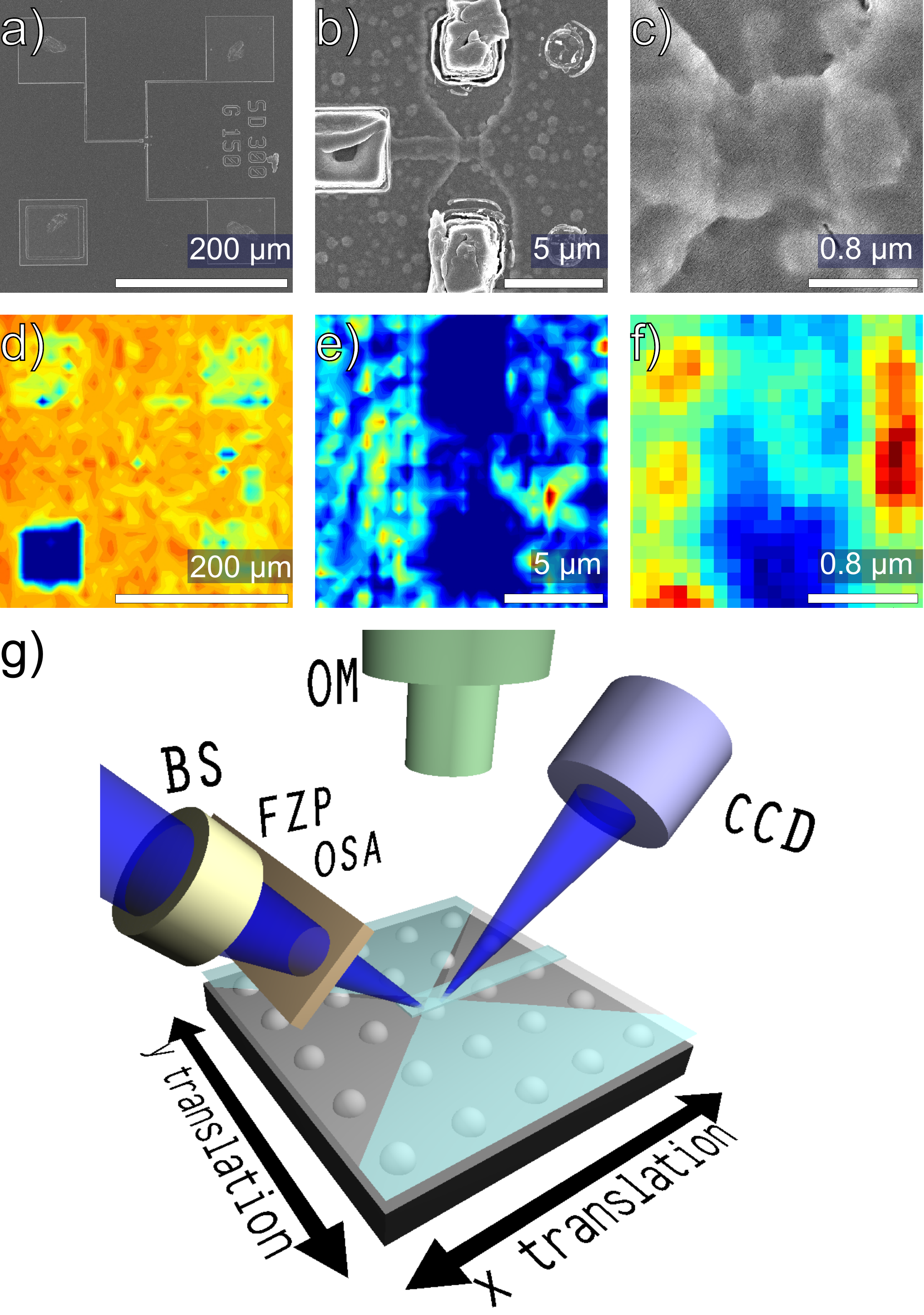}
\caption{(a)-(c) SEM images of the investigated dotFET device at different magnifications. The spotty surface is due to a thin Au layer to avoid charging during SEM. (d)-(f) according images obtained in scanning x-ray diffraction mode, with the same scaling. For (d) the diffractometer was tuned to the diffuse Si signal around (004), for (e) and (f) it was tuned to the SiGe dot signal at the (224) Bragg peak. (g) Schematics of the focusing setup (a) and the sample as it was mounted on the sample stage with respect to the beam direction. On the primary beam(PB) side, a set of a Fresnel zone plate (FZP), a beamstop (BS) and an order sorting aperture (OSA) was applied to focus the beam and eliminate higher harmonics. An optical microscope was mounted for a rough alignment of the sample.}
\label{fig:1}
\end{figure}

For the fabrication of the devices 4 inch Si(001) wafers were used with a sample layout fitting the requirements of island growth and later transistor processing. Within fields of 300$\mu$m x 300$\mu$m 2D arrays of pits with a period of 800 nm were defined by electron beam lithography (EBL) and reactive ion etching (RIE). Then 36 nm of Si as buffer layer were deposited at 450-550\textdegree C, followed by 6 monolayers (ML) of Germanium at 720\textdegree C resulting in dome-shaped islands with a diameter of 220 nm and a height of 45 nm (aspect ratio 0.2). The deposition of the 30 nm Si capping layer was done at a lower temperature of 360\textdegree C to avoid intermixing. The process flow was then continued by depositing a 400 nm thick SiO$_{2}$ isolation layer by plasma-enhanced chemical-vapor-deposition (PECVD) and removed by RIE in the active areas. A 15-nm-thick oxynitride-layer was grown as gate dielectric. Subsequently, the TiN/Al(1\% Si) gate layer was deposited by physical-vapor-deposition (PVD) at 350\textdegree C. The opened source/drain regions were implanted with As$^{+}$ ions at 5 keV, 10$^{15}$~cm$^{-2}$ dose. To recrystallize the damaged Si region, a single 25 ns-shot from a XeCl excimer laser ($\lambda$ = 308 nm) was applied, melting the amorphous volume. For the Al(1\% Si) gates wet etching was used, whereas the excess part of the TiN was removed by RIE. An 800-nm-thick isolation oxide was deposited by PECVD at 400\textdegree C, the contacts opened to the source and drain, followed by the PVD of a 905 nm thick Al(1\% Si) layer at 350\textdegree C. The metal was then removed from the surface except around the source/drain contacts and the contacts opened to the gate metal. A second thick layer of Al(1\% Si) was then deposited and patterned to leave metal lines connecting devices to the metal pads. Scanning electron microscopy (SEM) images of a fully-processed device are shown in Fig.~\ref{fig:1}. Panel (a) shows the layout of the device with contract pads (SEM), the region with the dotFET is enlarged in panels (b) and (c).  The electrical characterization of the processed dotFET devices compared with reference devices on the same wafers, processed outside of the regular dot arrays has confirmed the average increase of drain current between 20 and 60\%.\cite{Jovanovic2010}

\begin{figure}[tb]
\centering
\includegraphics[width=8.5cm]{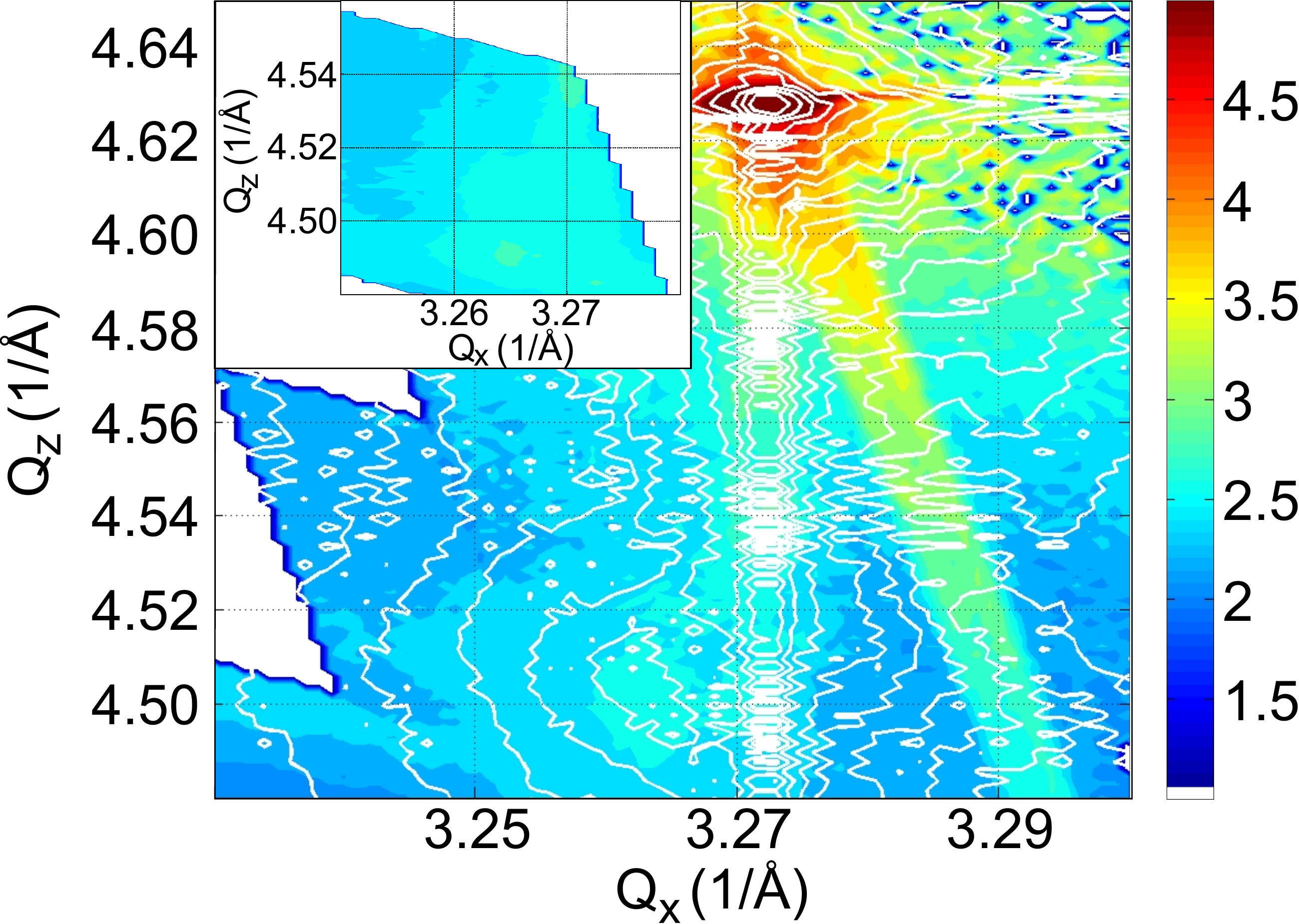}
\caption{Comparison of experimental data (colorplot) and a calculated reciprocal space map (contours) for the (224) Bragg Peak. In the inset a map for a neighboring dot is shown. The color bar shows the decadic logarithm of the intensity in counts per second.Q$_z$ and Q$_x$ denote reciprocal space coordinates along the [001] and [110] directions.}
\label{fig:2}
\end{figure}
%
X-ray experiments were carried out at beamline ID01 at the ESRF in Grenoble at an energy of 8 keV. The setup is sketched in Fig.~\ref{fig:1}(g): Fresnel zone plates (FZP) fabricated by Au-deposition on a lithographically structured Si support were used for focusing. The individual zones of the FZP with a thickness of about 1 $\mu$m introduce a phase shift of $\pi$ for the part of radiation traversing through the Au rings, thus acting as a phase grating.\cite{David2007} The resulting effective focus diameter, which is also influenced by beamline instabilities such as mutual vibrations of FZP and sample stage, was 400 nm FWHM.\cite{Diaz2009} At an incidence angle of 34.7\textdegree for the Si(004) or 79.6\textdegree for the (224) Bragg peak, the footprint on the sample is hence smaller than the period of the patterned SiGe dot array, which was 800 nm. This is true for the intense central part of the focused beam, especially at lower incidence angles a very small contribution to the scattered intensities may arise due to neighboring islands illuminated by the beam tails. 
Several difficulties are inherent to the diffraction experiment on a single island: first of all, the diffracted intensity is rather at the limit of detection due to the small island volume \cite{Stangl2009}. For the identification of the island beneath the transistor gate, the alignment has to be done using x-rays. Thus a characteristic signal has to be identified and the sample position scanned to determine the position of the x-ray focus on the sample. In order to locate the transistor the diffractometer angles were tuned to the diffuse scattering around the Si peak which yielded sufficient contrast as shown in Fig.~\ref{fig:1}(d). The metal contacts are clearly visible, and hence the center of the transistor can be located. In a second step, the goniometer was tuned to the scattering signal from the SiGe islands and by again mapping the intensity distribution in real space, the location of the islands was detected (Fig.~\ref{fig:1}(e,f)). Note that the center dots appear weaker in SXD, due to the thicker metalization layers on top of them. 
The reciprocal space map shown in Fig. \ref{fig:2} was recorded in several steps: after the position of the center dot was determined, the first part of the map covering the area of the (224) - SiGe signal in reciprocal space was recorded by performing a small omega scan (range 0.8\textdegree) while the detector was kept at a fixed position. Subsequently the detector arm was moved towards the substrate position in small steps, performing more omega scans to cover the rest of the space between the SiGe signal and the Si substrate peak. It turned out that the stability of the setup is sufficient to keep the dot within the incident beam during such a scan, resulting in a smooth signal from the dot. We additionally recorded a map of a neighboring island where it turned out that while the main position of the SiGe-signal stays roughly the same, more delicate features are different (see inset of Fig. \ref{fig:2}).
\begin{figure}[tb]
\centering
\includegraphics[width=8cm]{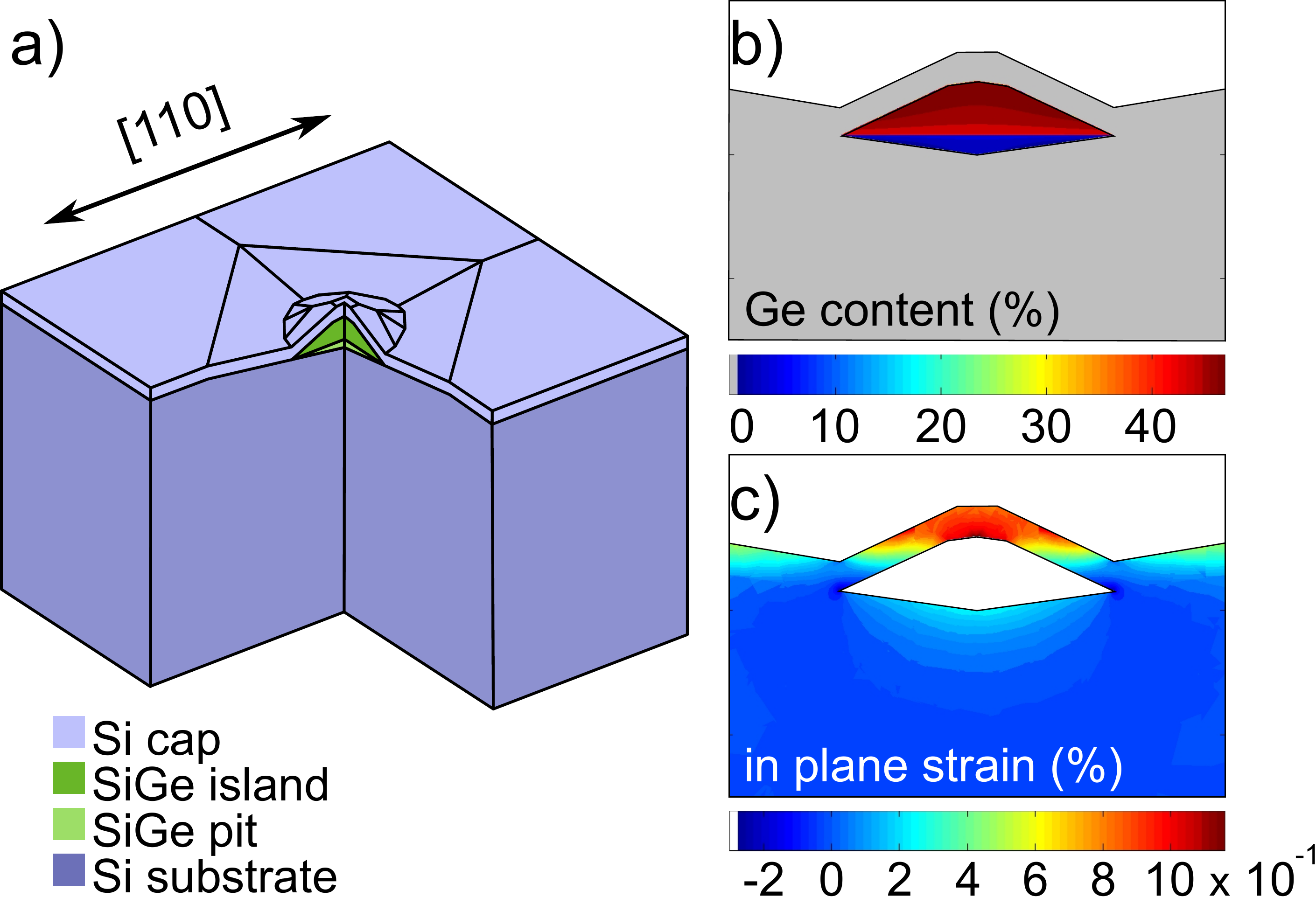}
\caption{(a) Sketch of the epitaxially grown parts represented in the model geometry used for FEM calculations with a faceted island in a pit, covered by a 24 nm Si cap layer. (b) and (c) show 2D maps of the Ge concentration and the in-plane strain, respectively. For clarity, the strain field is shown only for the Si cap and substrate. Both maps are oriented along the [110] direction.}
\label{fig:3}
\end{figure}

To quantify the strain in the Si bridge above the island in the transistor, we used FEM model calculations and simulations of the x-ray intensity distribution based on kinematical diffraction theory. \cite{Pietsch2004} A simplified model geometry was used, containing the epitaxially grown components of the structure, namely the SiGe dot and the Si cap as well as the lowermost components of the gate stack (nitride layers and Al-gatefinger) which directly contribute to the strain state of the buried structure. The geometry for the FEM calculation was derived from an AFM analysis of uncapped and capped samples grown under the same conditions (see Fig. \ref{fig:3}a) and is also in good agreement with TEM images \cite{Jovanovic2010}.
Resonant Raman spectroscopy experiments on buried SiGe islands grown on \textit{flat} Si substrates revealed a considerable dependence of the strain on the Si cap-layer thickness.\cite{Bonera2009} From the x-ray intensity oscillation period along the crystal truncation rod, the thickness of the Si capping layer was determined to be only about 24~nm, instead of initially 30~nm. This reduced thickness was used for the FEM calculations. From selective etching experiments and previous x-ray and TEM studies, the Ge content in the bottom part of the pit is known to be very low (about 5\%), increases sharply towards the upper dome-shaped part, where it varies rather gradually. In the model, the downward looking part of the island in the pit was therefore ''filled'' with a SiGe compound with a maximum Ge content of 5 \%. On the upper dome shaped island section a three dimensional gradient was applied following a square root function in vertical direction and also laterally decreasing. This combination results in a realistic onionskin-like Ge distribution (see Fig. \ref{fig:3}b). Using these functions, the absolute Ge content values as well as the initial strains in the gate stack are fitted, resulting in a Ge content of 43\% at the bottom and 48\% at the top of the dome-shaped part of the island, resulting in an average Ge content of about 40 \% for the whole SiGe structure. A tensile strain of 0.007 was applied to nitride sections of the gate stack, which has only a minor influence on the strain state within the Si channel. The fit between x-ray simulations based on a model including the gate stack and the measured RSM can be seen in Fig. \ref{fig:2}. With these simulations, also the strain distribution within the SiGe island, and the Si bridge are obtained. In the 24~nm thick capping layer, maximum tensile strains of about 1 \% are achieved, the values in the main active region of the Si channel are above 0.8\% (see Fig. \ref{fig:3}c).

In conclusion, we have used x-ray diffraction with a beam focused to 400~nm diameter to determine the strain state of the Si n-channel above a buried SiGe island, within a fully functioning field effect transistor. Tensile strain values up to 1\% can be achieved using this approach, i.e., within a fully pseudomorphic structure without defects.

The authors thank the entire crew at ID01 for their excellent support and B. Mandl for the SEM images. This work was supported by the EC d-DOTFET project (012150-2) and the FWF Vienna (SFB025 IR-On).

%
\end{document}